\documentclass[twocolumn,pra,superscriptaddress,nofootinbib]{revtex4}

\usepackage{graphicx}
\usepackage{changes}
\usepackage{amssymb}
\usepackage{algpseudocode}
\usepackage{amscd}
\usepackage{algorithm}

\usepackage{tikz}
\usetikzlibrary{chains}
\usetikzlibrary{fit}
\usepackage{xcolor}
\usepackage{epsfig}

\usepackage{amsthm}
\usepackage{amsmath}
\usepackage{amsfonts}
\usepackage{amssymb,amstext}

\usepackage[colorlinks=true,urlcolor=blue, hyperindex] {hyperref}

\usepackage{epigraph}

\usepackage{etoolbox}

\makeatletter
\patchcmd{\epigraph}{\@epitext{#1}}{\itshape\@epitext{#1}}{}{}
\makeatother

\theoremstyle{plain}

\theoremstyle{definition}
\newtheorem{mydef}{Definition}

\newcommand{\id}{\mathrm{id}}
\newcommand{\ket}[1]{\left| #1 \right \rangle}
\newcommand{\bra}[1]{\left \langle #1 \right|}

\newcommand*{\cD}{\mathcal{D}}
\newcommand*{\cI}{\mathcal{I}}
\newcommand*{\cM}{\mathcal{M}}
\newcommand*{\cT}{\mathcal{T}}

\newcommand*{\cU}{\mathcal{U}}

\DeclareMathOperator{\tr}{tr}

\definecolor{nblue}{rgb}{0.2,0.2,0.8}
\definecolor{ngreen}{rgb}{0.2,0.7,0.2}
\definecolor{nred}{rgb}{0.8,0.2,0.2}
\definecolor{nblack}{rgb}{0,0,0}

\definecolor{darkgreen}{rgb}{0,0.4,0}
\definecolor{orange}{rgb}{1,0.5,0}

\newcommand{\proj}[1]{\ket{#1}\!\!\bra{#1}}
\newcommand{\ketbra}[2]{\ket{#1}\!\!\bra{#2}}

\begin{document}

\title{Quantum clocks and their synchronisation ---  the Alternate Ticks Game}
\author{Sandra Rankovi\'{c}}
\affiliation
{Institute for Theoretical Physics, ETH Zurich, 8093 Zurich, Switzerland}
\author{Yeong-Cherng Liang}
\affiliation{Department of Physics, National Cheng Kung University, Tainan 701, Taiwan}
\affiliation
{Institute for Theoretical Physics, ETH Zurich, 8093 Zurich, Switzerland}
\author{Renato Renner}
\affiliation
{Institute for Theoretical Physics, ETH Zurich, 8093 Zurich, Switzerland}

\begin{abstract} 
Time plays a crucial role in the intuitive understanding of the world around us. Within quantum mechanics, however, time is not usually treated as an observable quantity; it enters merely as a parameter in the laws of motion of physical systems. Here we take an operational approach to time. Towards this goal we  consider quantum clocks, i.e., quantum systems that generate an observable time scale. We then study the quality of quantum clocks in terms of their ability to stay synchronised. To quantify this, we introduce the ``Alternate Ticks Game'' and analyse a few strategies pertinent to this game.
\end{abstract}


\maketitle

\epigraph{``He who made eternity out of years remains beyond our reach. His ways
remain inscrutable because He not only plays dice with matter but also with time."}{Karel V. Kucha\v{r}~\cite{Kuchar}}

\section{Introduction}

{\it Time} is a central concept used for the description of the world around us. We perceive it intuitively and we can measure it to very good precision \cite{Wineland, Haroche, AtomicClocks}. Nevertheless,  it remains as one of the biggest unknowns of modern physics~\cite{EinsteinClock, Heisenberg, PauliTwo, DeWitt,  Peres, Busch, Maccone, Kuchar}. In quantum mechanics, time is not considered an observable, but plays merely a parametric role in the equation of motion. In fact, it cannot be treated like other observables, as there cannot exist a self-adjoint time operator that is canonically conjugate to a Hamiltonian having a semi-bounded  spectrum~\cite{PauliTwo}. There have been various attempts, though, to establish an understanding of time that goes beyond this parametric view~\cite{AharonovBohm,PageWootters, Wootters, Miyake, Vlatko, LLoydQuantumTime,ThermodynamicsOfTime}, e.g., by  considering time as arising from correlations between  physical systems. In this work we intend to take further steps in this direction. 

Our approach is operational, in the sense that we study clocks, i.e., physical systems that provide time information. This is motivated by earlier work~\cite{TimeBook,Wootters}, where it has also been  suggested to distinguish between {\it coordinate time} and {\it clock time}. While the first refers to a parameter used within a physical theory, the latter is an observable quantity. Here we are interested in the latter. Specifically, we view clock time as the observable output of a quantum system, hereafter  referred to as a {\it quantum clock}. 

Clearly, it is desirable to have a notion of quantum clocks that agrees with our usual perception of classical clocks in a certain limit. Yet, for our operational approach towards time, we shall assume only some basic features that a  clock must exhibit.  Importantly, a clock should generate a clock time that can be used to order events. This feature will play a central role in our treatment and will be captured below by our definition of a {\it time scale}.

Within Newtonian theory, there do not seem to be any fundamental limitations to the accuracy  to which  clocks can stay synchronised (see e.g.,~\cite{Wineland,AtomicClocks}).  This gives rise to a {\it global} notion of time. A  notion of global time is also assumed in the usual treatment of quantum mechanics. However, the corresponding global structures have various non-trivial features, which are a topic of ongoing research (see, e.g.,~\cite{Oreshkov, Bancal:2012aa, Brukner, Curchod:2014, Oreshkov2014, Amin}). Furthermore,  a global notion of time is generally not achievable  in  relativistic theories~\cite{DeWitt, UnruhTime, Isham, Anderson, Kuchar, PhilosophyTime}. In our operational approach, we will thus treat quantum clocks as {\it local} physical systems, governed by the laws of quantum mechanics.   This raises the question of {\it synchronisation}. In operational terms, we are interested to know whether  two clocks that are separated, so that no communication between them is possible, can order events consistently. If this is the case, we say that the time scales generated by the two clocks are compatible. We note that this question relates to the  general idea of reference frames in quantum theory~\cite{Spekkens} and, in particular, the question of how well such reference frames can be correlated. 

To quantify the level of synchronisation between clocks, we consider a particular scenario, which we term the {\it Alternate Ticks Game}. In this game, two players, each equipped with a quantum clock, are asked to send {\it tick signals} in alternating order to a referee.  The number of tick signals that they can generate in the correct order is then taken as a measure of how well the clocks are synchronised and, hence, of the quality of these clocks. 

Quantum clocks should, by definition, produce a measurable output --- the clock time. As any measurement of a quantum system unavoidably disturbs its state, one should expect fundamental limitations to the precision of quantum clocks, and hence the level of synchronisation attainable between them. It is one of the major motivations of the present work to understand these limitations. 

This paper is structured as follows. In Section~\ref{sec:Time}, we introduce our definition of a {\it  quantum clock} and the associated {\it  time scale}. Section~\ref{sec:Cont}  is concerned with a {\it continuity} condition, which ensures that quantum clocks can be regarded as self-contained devices, in particular that they do not implicitly rely on a time-dependent control mechanism. Then, in Section~\ref{sec:Game}, we consider the synchronisation of clocks and introduce the {\it  Alternate Ticks Game}, which serves as a method to investigate the quality of quantum clocks. The numerical results of some possible strategies pertinent to this game are presented in Section~\ref{sec:Results}. In our conclusions in Section~\ref{sec:Conclusion} we discuss possible future research directions that are spurred by this work.

\section{Quantum clocks and time scales}
 \label{sec:Time}

\begin{figure}[h]
\begin{tikzpicture}
\draw (4,1.7) ellipse(0.55cm and 0.65cm) node {$C$};
\draw [<->] (4,0.9) -- (4,0.35);
\draw (3.6,-0.6) rectangle(4.35,0.05);
\draw (4,-0.3) node {$T_{i}$};
\draw (4.9,0.7) node {$\mathcal{M}_{C\to CT}$};
\end{tikzpicture}
\caption{{\it Model of a quantum clock.} A  quantum clock consists of a clockwork $C$ that interacts with tick registers $T_1, T_2, \ldots$. These registers are propagated away from $C$ after their interaction, keeping a record of the clock time.  }
\label{fig:Fig1}
\end{figure}
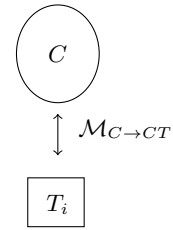

The aim of this section is to motivate and define our notion of a {\it quantum clock} (Definition~\ref{Def1}) and the derived concept of a {\it time scale} (Definition~\ref{Def2}).  Roughly speaking, a quantum clock is a quantum physical system equipped with a  mechanism that generates time information.  We model this mechanism as a process involving the following two systems (see Fig.~\ref{fig:Fig1}). 
\begin{itemize}
\item {\it Clockwork (denoted $C$):} This is the dynamical part of the clock whose state evolves with respect to the coordinate time.  In addition, it interacts with its environment, thereby outputting time information.  

\item {\it Tick registers (denoted $T_1, T_2, \ldots $):}  These belong to the environment of the clockwork.  Each tick register $T_i$ is briefly in contact with the clockwork and records the output of the latter. After this interaction, it is separated from the clockwork, keeping a record of the clock time. 
\end{itemize}

The  evolution of a physical clock appears to be continuous. However, as the notion of continuity implicitly refers to some underlying time parameter, we model the evolution of a quantum clock more generally as a sequence of discrete steps. Continuity may then be approximated by imposing an additional condition (see Definition~\ref{Def3} below). Each of the discrete evolution steps consists of the interaction between the clockwork $C$ and a fresh tick register $T_i$. Crucially, all steps are governed by the same dynamics, specified by a map $\cM$. Accordingly, all tick registers $T_i$ are taken to be isomorphic to a virtual system~$T$.  Furthermore, we specify the initial state of the clockwork $C$ by a density operator, denoted by $\rho^0_C$.  This idea is captured by the following definition. 

 \begin{mydef} \label{Def1}
  A {\it quantum clock}  is defined by a pair $(\rho^0_C, \cM_{C \to CT})$, where $\rho^0_C$ is a density operator of a system $C$, called {\it clockwork}, and $\cM_{C \to C T}$ is a  completely positive trace-preserving (CPTP) map from $C$ to a composite system $C \otimes T$. 
\end{mydef}

\begin{figure}[h]
\includegraphics[scale=0.35]{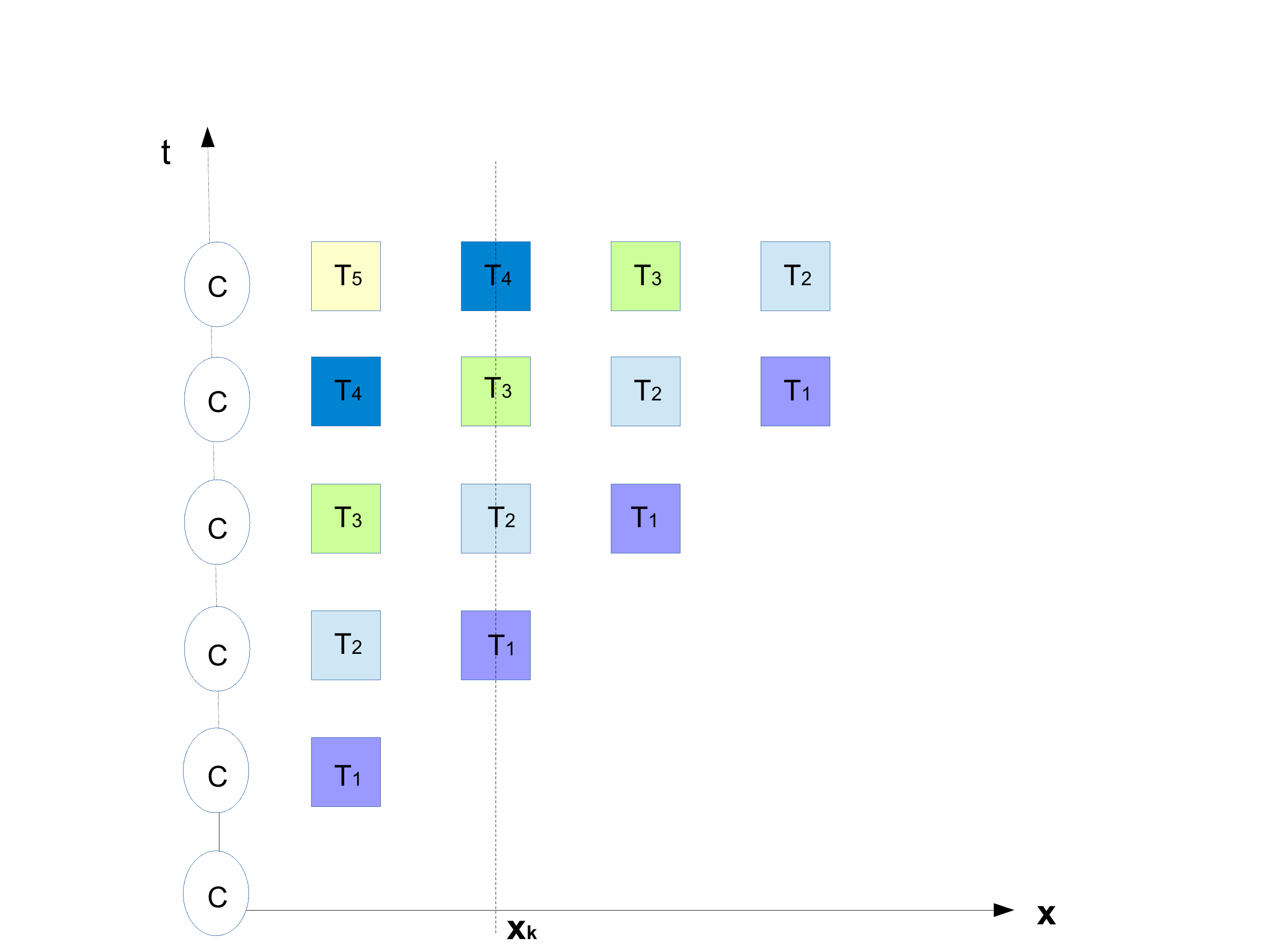}  
\caption{{\it Tick registers and the time scale.} Intuitively, one may think of the tick registers $T_i$ as propagating through space (coordinate~$x$), starting from the location of the clockwork $C$, in terms of coordinate time  (denoted by~$t$). An observer located at position $x_k$ would observe them in sequential order (with respect to~$t$). The content of the sequence of tick registers defines a time scale. }  
\label{fig:Fig2}
\end{figure}

The clock time produced by a clock gives rise to a {\it time scale}. Since, in our model, the clock time is recorded by the tick registers, a time scale corresponds to the cumulative content of these registers (see Fig.~\ref{fig:Fig2}).  Formally, this is defined as follows. 

\begin{mydef} \label{Def2}
  Let $(\rho^0_C, \cM_{C \to CT})$ be a quantum clock and let $T_1, T_2, \ldots,$ be a sequence of systems isomorphic to $T$, called {\it tick registers}. Then the  {\it time scale} generated by the quantum clock is the sequence $\{\rho^N_{T_1 \cdots T_N}\}_{N \in \mathbb{N}}$ of density operators on $T_1 \otimes \cdots \otimes T_N$ defined by the partial trace over $C$ of 
   \begin{align} \label{eq_rhoNdef}
   \rho^N_{C T_1 \cdots T_N} = \bigcirc_{j = 1}^N \cM_{C \to C T_j}(\rho^0_C) \ ,
 \end{align}
 where $\cM_{C \to C T_j}$ denotes the completely positive map that acts like $\cM_{C \to C T}$ from $C$ to $C \otimes T_j$, while leaving $T_1, \ldots, T_{j-1}$ unchanged.
  \end{mydef}
  
It is easy to verify that for any $N' > N$ we have
  \begin{align}
    \rho^N_{T_1 \cdots T_N} = \tr_{T_{N+1} \cdots T_{N'}}(\rho^{N'}_{T_1 \cdots T_{N'}}) \ .
  \end{align}
  The sequence $\{\rho^N_{T_1 \cdots T_N}\}_{N \in \mathbb{N}}$ of density operators is thus just a way to specify the state of the system\footnote{Note that the tensor product of infinitely many finite Hilbert spaces is not necessarily a separable Hilbert space.} $T_1 \otimes T_2 {\otimes \cdots}$ formed by all (infinitely many) tick registers after infinitely many invocations of the map~$\cM$. 
  
Realistic clocks consist of various physical components, such as a power source to drive the clock.  Our definition of a quantum clock,  i.e., the pair $(\rho^0_C, \cM_{C \to C T})$, does not specify these explicitly. It rather provides an abstract description of how the clock generates time information. 
While such an abstract model is suitable for our considerations,  one should keep in mind that it is quite generic. That is, any physically realisable clock admits an abstract description in terms of a pair  $(\rho^0_C, \cM_{C \to CT})$, but not any such pair corresponds to a physically realisable quantum clock. Indeed, in Section~\ref{sec:Cont} we will  introduce an additional condition that accounts for certain physical constraints.

Our definition of a quantum clock also does not explicitly require certain features that may be expected from a  clock --- constant time intervals, cyclicity, and unidirectional evolution. These turn out to be unnecessary  for the operational task of synchronising quantum clocks. Take, as an example, the notion of constant time intervals. Intuitively, this seems to be necessary to  determine the duration of events in some well-defined units (e.g., seconds). But, if the duration of a second were to change every time the clock should tick (i.e., if it was not a constant), and if {\it  all} clocks would perform these adjustments automatically without us noticing, then such a change would not lead to any observable consequences.

In the following we will often consider particular constructions of quantum clocks where the map $\cM$ can be written as the concatenation of two CPTP maps, 
\begin{align} \label{eq_typicalclock}
  \cM_{C \to C T} = \cM^{\mathrm{meas}}_{C \to C T} \circ \cM^{\mathrm{int}}_{C \to C} \ .
\end{align}
$\cM^{\mathrm{int}}$ acts only on the clockwork $C$ and $\cM^{\mathrm{meas}}$ corresponds to a  measurement on $C$ of the form
\begin{align} \label{eq_meas}
  \cM^{\mathrm{meas}}_{C \to C T} : \, \rho_C \mapsto \sum_{t \in \cT} \sqrt{\pi_t} \rho_C \sqrt{\pi_t} \otimes \proj{t}_T \ ,
\end{align}
where $\{\pi_t\}_{t \in \cT}$ is a Positive-Operator Valued Measure (POVM)\footnote{A POVM is a family of positive semidefinite operators $\pi_t$ such that $\sum_t \pi_t =  \id$, where $\id$ denotes the  identity operator.}  on $C$ and where $\{\ket{t}\}_{t \in \cT}$ is a family of orthonormal vectors on the Hilbert space of $T$. $\cM^{\mathrm{int}}$ can be interpreted as the internal mechanism that drives the clockwork, whereas $\cM^{\mathrm{meas}}$ extracts  time information and copies it to the tick registers.
 
For a simple example of a clock of the form~\eqref{eq_typicalclock}, assume that the clockwork $C$ is binary (i.e., described by the Hilbert space $\mathbb{C}^2$), with orthonormal states $\ket{0}$ and $\ket{1}$, and that the map $\cM^{\mathrm{int}}$ flips the state on $C$, corresponding to a logical NOT operation. Furthermore, $\cM^{\mathrm{meas}}$ could be a measurement defined by the projectors $\pi_0 = \proj{0}$ and $\pi_1 = \proj{1}$.  Then, with $C$ initially set to $\ket{0}$, the resulting time scale $\{\rho_{T_1 \cdots T_N}^N\}_{N}$ would consist of operators of the form
 \begin{align*}
   \rho_{T_1 \cdots T_N}^N = \proj{1} \otimes \proj{0} \otimes \proj{1} \otimes \proj{0} \otimes \proj{1} \otimes \cdots  .
 \end{align*}
 Because of its deterministic character such a clock would be perfectly precise, in the sense that two such clocks could stay synchronised infinitely long. In particular, they would perform arbitrarily well in the Alternate Ticks Game defined in Section~\ref{sec:Game}.

\section{Continuous quantum clocks} \label{sec:Cont}

It appears to be easy to physically realise the simple clock described in the previous section  --- one merely needs to implement a NOT gate and a measurement. However, it is unclear whether these operations can be carried out without the help of an additional {\it time-dependent} control mechanism, i.e., one that turns on and off certain interactions at well-defined points in coordinate time. And if such a control mechanism is necessary, the clock can of course no longer be regarded as a self-contained device.  This motivates an additional assumption, namely that the clockwork evolves continuously.   As our model is inherently discrete, we formalise this by the requirement that the individual steps of the evolution can be made arbitrarily small.

 \begin{mydef}
    \label{Def3}
    A  quantum clock $(\rho^0_C, \cM_{C \to C T})$ is called  {\it $\epsilon$-continuous} if the map $\cM_{C \to C T}$ restricted to $C$ is $\epsilon$-close to the identity map $\cI_C$, i.e., 
    \begin{align}
      \| \tr_{T} \circ \cM_{C \to C T} - \cI_C \|_{\diamond} \leq \epsilon
    \end{align}
    where $\| \cdot \|_{\diamond}$ is the diamond norm. 
 \end{mydef}

$\epsilon$-continuity, for arbitrarily small values~$\epsilon$, is a necessary condition for a  clock to be self contained.\footnote{That a clock is self contained means that any possible control mechanism is regarded as part of it.  In this case the Hamiltonian of the system is time-independent. Considering small steps in coordinate time, such a clock can always be modelled as an $\epsilon$-continuous quantum clock, for any $\epsilon> 0$.}  
Furthermore, combined with our requirement that all evolution steps are identical it ensures that the evolution of the state of the clockwork does not have an implicit time dependence. In particular, it does not depend on the timing of the interaction between the clockwork and the individual tick registers. To see this, consider a  gear system that, in each step, provides a fresh tick register~$T$ in a fixed state $\sigma_T$ and then lets the joint state of the clockwork~$C$ and the tick register~$T$ evolve according to some map $\cU = \cU_{C T}$. This corresponds to a quantum clock defined by the map
\begin{align} \label{eq:Umap}
  \cM_{C \to CT} : \, \rho_C \mapsto \cU(\rho_C \otimes \sigma_T) \ .
\end{align}
In particular, if the initial state of the clockwork~$C$ is $\rho_C^0$ then the states after the first and the second step are
\begin{align*}
  \rho^1_{C} & = (\tr_T  \circ\, \cU) (\rho^0_C \otimes \sigma_T)  \\
    \rho^2_{C} & = (\tr_T \circ\, \cU)(\rho^1_C \otimes \sigma_T) \ .
\end{align*}
This may now be compared to a situation where the gear system fails to deliver a fresh tick register between the first and the second execution of $\cU$. In this case the state of the clockwork after the second step would be
\begin{align*}
  \bar{\rho}^2_{C} = (\tr_T \circ\, \cU \circ \cU)(\rho^0_C \otimes \sigma_T) \ .
\end{align*}
For any map $\cM$ that is $\epsilon$-continuous one can choose the corresponding map $\cU$ such that  $\cU = \cI + \cD$ with $\| \cD \|_{\diamond} \leq \epsilon$. Inserting this into the above expressions for the states gives
\begin{align*}
  \rho_C^2 & =  \rho^0_C + 2(\tr_T \circ \cD)(\rho^0_C \otimes \sigma_T) +\delta_C \\
    \bar{\rho}_C^2 & =  \rho^0_C + 2(\tr_T \circ \cD)(\rho^0_C \otimes \sigma_T) +  \bar{\delta}_C 
\end{align*}
with 
\begin{align*}
  \delta_C & = (\tr_T \circ \cD)\bigl((\tr_T \circ \cD)(\rho^0_C \otimes \sigma_T) \otimes \sigma_T\bigr) \\
  \bar{\delta}_C & = (\tr_T \circ \cD \circ \cD)(\rho^0_C \otimes \sigma_T) \ .
\end{align*}
Using the fact that both the partial trace $\tr_T$ and the tensoring map $X_C \mapsto X_C \otimes \sigma_T$ have diamond norm equal to~$1$, we find  
\begin{align*}
  \| \delta_C \|_1 \leq   \| \cD \|_{\diamond}^2 \leq \epsilon^2 \quad \text{and} \quad
  \| \bar{\delta}_C \|_1  \leq    \| \cD \|_{\diamond}^2 \leq \epsilon^2 \ .
\end{align*}
From this we obtain a bound on the distance between the two states, 
\begin{align*}
\| \rho_C^2- \bar{\rho}_C^2  \|_1 
= \| \delta_C - \bar{\delta}_C \|_1
\leq \| \delta_C \|_1 + \|\bar{\delta}_C \|_1 \leq 2 \epsilon^2 \ .
\end{align*}

Suppose now that the clock runs for $N+1$ steps where, as above, in each step the state of $C \otimes T$ undergoes a mapping $\cU$. Let $\rho^{N+1}_C$ be the state of $C$ under the assumption that a fresh tick register (initialised in state $\sigma_T$) is provided in between any two executions of $\cU$. Let furthermore $\bar{\rho}^{N+1}_C$ be the corresponding state where this condition may fail with probability~$p$ in between any of the steps. Generalising the reasoning above, we find that the distance between the two states is bounded by
\begin{align*}
  \| \rho_C^{N+1} - \bar{\rho}_C^{N+1} \|_1 \leq 2 N p \epsilon^2 \ .
\end{align*}
Note that the value $N$ necessary to achieve a certain change of the state of the clockwork scales inverse proportionally to $\epsilon$, i.e., $N = c / \epsilon$ for some constant~$c$ (independent of~$\epsilon$). Inserting this in the above bound we conclude that the distance is of the order
\begin{align*}
  \| \rho_C^{N+1} - \bar{\rho}_C^{N+1} \|_1 \leq O(\epsilon)\  ,
\end{align*}
i.e.,  the effect onto the state of the clockwork~$C$ due to failures in the replacement of the tick registers disappears as $\epsilon$ tends to~$0$. In other words, the performance of a continuous quantum clock does not depend on the exact timing of the insertion of the tick registers. 
 
The clock with the NOT-based clockwork described at the end of Section~\ref{sec:Time} is not $\epsilon$-continuous for any $0 \leq \epsilon < 2$, as in each step the state of $C$ is deterministically changed to an orthogonal one. In fact, constructing a continuous  clock is a bit more challenging. One conceivable approach could be to split the NOT operation used in the construction into small identical steps. Specifically, the map $\cM^{\mathrm{int}}$ could be defined as an $n^{\mathrm{th}}$ root of the NOT operation, for some large $n$, corresponding to a rotation in state space by a small angle. For any $\epsilon > 0$ and sufficiently large $n$ this map would be $\epsilon$-close to the identity. However, it is unclear how to choose the measurement $\cM^{\mathrm{meas}}$. In fact, if the described small rotations are applied in between two measurements, its outcome would no longer be deterministic.  Consequently, the clock would  generate a rather randomised time scale and thus not be very precise. 
 
As we shall see, the price to pay for more precision seems to be an increased size of the clockwork~$C$. To illustrate this, we consider a clockwork that mimics the propagation of a wavepacket. Specifically, let $C$ be equipped with an orthonormal basis $\{\ket{c}\}_{c = 0, \ldots, d-1}$,  for $d \in \mathbb{N}$ and define, for any $\bar{c} \,\in\{0, \ldots, d-1\}$ and $\Delta \ll d$,
\begin{align}
   \Pi_{\bar{c}} = \sum_{c: \, |c - \bar{c}| \leq \Delta} \proj{c} \ .
\end{align}
Furthermore, assume that the initial state $\rho^0_C$ is contained in the support of the projector  $\Pi_{c_0}$ with $c_0 = 0$, i.e., 
\begin{align}
  \tr(\Pi_{c_0} \rho_C^0) = 1 \ .
\end{align}
Intuitively, one may think of $\rho_C^0$ as a wavepacket of breadth~$\Delta$ localised around $c_0$.  Let $\cM^{\mathrm{int}}$ be a map that moves this wavepacket by some fixed distance  $\nu \ll 1$, in the sense that the state $\rho_C^N$ of the clockwork after $N$ applications of $\cM^{\mathrm{int}}$ [see~\eqref{eq_rhoNdef}] satisfies
\begin{align} \label{eq_regularmovement}
  \tr(\Pi_{c_N} \rho_C^N) \approx 1 \quad \text{for $c_N = \lfloor N \nu \rfloor$} 
\end{align}
(provided that $N \nu + \Delta < d-1$).  Finally, the measurement  $\cM^{\mathrm{meas}}$ [see~\eqref{eq_meas}] could be defined by the POVM elements 
\begin{align}\label{Eq:Def:pi}
  \pi_0 = \id_C - \delta \Pi_{\bar{c}}  \quad \text{and} \quad \pi_1 =  \delta \Pi_{\bar{c}} 
\end{align}
for $\bar{c}= d-1$ and some $\delta \ll 1$. It is easy to see that, for any $\epsilon > 0$, the clock can be made $\epsilon$-continuous by choosing $\nu$ and $\delta$ sufficiently small.  

We note that this construction relies on the assumption that the  (mimicked) wavepacket propagates regularly (such that condition~\eqref{eq_regularmovement} is satisfied). This is however only possible if its breadth~$\Delta$ is sufficiently large so that the wavepacket has not too much spread in momentum. Since the dimension~$d$ of $C$ must certainly be larger than $\Delta$, the precision of the proposed construction depends strongly on the size of the clockwork. 

To analyse the time scale that the clock generates we first observe that, as long as $N \nu + 2 \Delta < d$, the state of the clockwork $\rho_C^N$ has only negligible overlap with $\pi_1$ and is therefore not disturbed by the measurement $\cM^{\mathrm{meas}}$. During this phase, the clock would output a series of states $\ket{0}$ to the tick registers. Only once $N \nu$ is approximately equal to $d$, the clock would at some point output a state $\ket{1}$.  Hence, assuming that $\Delta \ll d$, the first ``tick'' almost never occurs before  $N_{\min} = d/\nu$ steps, but is very likely to occur before $N_{\max} = d/\nu + O(1/\delta)$ steps.  Since the length of the interval $[N_{\min}, N_{\max}]$ can be made short relative to $N_{\min}$ by choosing $d \gg \nu / \delta$, the clock produces a relatively precise first tick. However, after this first tick, the clock would in each step output  $\ket{1}$ with probability~$\delta$, resulting in a rather randomised pattern of ticks. 

The construction above could be leveraged to a more useful clock by adapting the measurement $\cM^{\mathrm{meas}}$ such that it resets the state of $C$ to $\rho^0_C$ whenever the output $\ket{1}$ is written to the time register. Formally, this corresponds to a clock where the CPTP map defined by Eq.~\eqref{eq_meas} is replaced by 
\begin{equation}\label{eq_meas_reset}
\begin{split}
  \cM^{\mathrm{meas}}_{C \to C T} : \, \rho_C \mapsto &\sqrt{\pi_0}\, \rho_C\, \sqrt{\pi_0} \otimes \proj{0}_T \\
  &+\tr (\pi_1\, \rho_C)\, \hat{\rho}_C  \otimes \proj{1}_T  
\end{split}
\end{equation}
with $\hat{\rho}_C = \rho^0_C$. The resulting time scale would then consist of almost equally long sequences of states $\ket{0}$ separated by states $\ket{1}$. However, the precision depends crucially on the size $d$ of the clockwork (see Section~\ref{sec:Results} for numerical results). In fact, it appears to be impossible to construct continuous clocks of finite size that are infinitely precise.

 \section{Synchronisation of clocks --- the Alternate Ticks Game}  \label{sec:Game}
 
So far we have  specified what type of physical systems we consider as clocks, but we have not yet provided any criteria to assess their capacity to generate precise time information. As it seems to be impossible to define the accuracy of a clock in an operational manner without comparing it to another one, we consider multiple clocks and ask how well they can stay synchronised. 

There are various ways to define {\it synchronisation} between two clocks. In general, synchronisation means that the time scales generated by the two clocks are in some sense compatible.  The strongest possible criterion would be that they are identical. However, such a perfect synchronisation would require perfectly precise clocks, which cannot be achieved with continuous clocks of finite size.

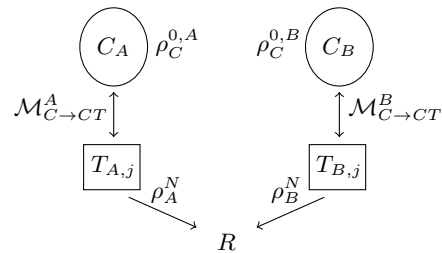
\begin{figure}[h]
\begin{tikzpicture}
\draw (3.1,1.5) node {$\mathcal{M}^A_{C\to CT}$};
\draw (7.55,1.5) node {$\mathcal{M}^B_{C\to CT}$};
\draw (4.65,2.35) node {$\rho^{0,A}_C$};
\draw (6.0,2.35) node {$\rho^{0,B}_C$};
\draw (3.8,2.3) ellipse(0.45cm and 0.52cm)   node {$C_A$};
\draw [<->] (3.8,1.7) -- (3.8,1.1);
\draw [<->] (6.8,1.7) -- (6.8,1.1);
\draw (4.15,0.4) rectangle(3.4,1);
\draw (3.8,0.7) node {$T_{A,j}$};
\draw (7.15,0.4) rectangle(6.4,1);
\draw (6.8,0.7) node {$T_{B,j}$};
\draw (6.8,2.3) ellipse(0.45cm and 0.52cm) node {$C_B$};
\draw [->] (4.0,0.3) -- (4.9,-0.1);
\draw [->] (6.6,0.3) -- (5.7,-0.1);
\draw (4.5,0.4) node {$\rho_A^N$};
\draw (6.1,0.4) node {$\rho_B^N$};
\draw (5.3,-0.3) node {$R$};
\end{tikzpicture}
\caption{{\it Schematic representation of the Alternate Ticks Game.} Two players $A$ and $B$ choose quantum clocks, possibly under certain constraints, e.g., a bound on the dimension of their clockworks, $C_A$ and $C_B$. The clocks are defined by initial states $\rho^{0,A}_C$, $\rho^{0,B}_C$ as well as maps $\mathcal{M}^A_{C\to CT}$, $\mathcal{M}^B_{C\to CT}$,
 respectively. Each of them generates a stream of tick registers, denoted by $T^A_{j}$ and $T^B_{j}$, whose cumulative contents are denoted by $\rho^N_A$ and $\rho^N_B$, respectively. The tick registers are sent to a referee~$R$, who checks the alternate ticks condition.}
\label{fig:Fig3}
\end{figure}

In the following, we  introduce a quantitative measure for synchronisation. We formulate it in terms of a game, the {\it Alternate Ticks Game}. It involves two collaborating players, $A$ and $B$, each of them equipped with a quantum clock.  The players can agree on a common strategy, but are not allowed to communicate once the game has begun. They are asked to provide {\it ticks} to a referee, who checks whether these ticks are received in an alternating order --- first from $A$, then from $B$, then again from $A$, and so on (see Fig.~\ref{fig:Fig3}).   The goal of the players is to maximise the number of  ticks respecting the posed alternate ticks condition.  

Formally, the strategy of the players is defined by their respective choice of a quantum clock, $(\rho^{0,A}_C, \cM^A_{C \to C T})$ and $(\rho^{0,B}_C, \cM^B_{C \to C T})$  (cf.\ Def.~\ref{Def1}). This choice may be subject to certain constraints, e.g., on the size of the clockwork. To capture the idea that the players have no access to additional time information, we do not allow them to carry out any non-trivial operations on the individual tick registers $T_j$ (such as operations that depend on $j$). Specifically, we shall assume that they simply transmit their tick registers to the referee. The referee continuously monitors the incoming stream of tick registers from both players via predefined projective measurements $\{\tau^A, \id_T - \tau^A\}$ and $\{\tau^B, \id_T - \tau^B\}$,  whose outcomes are interpreted as ``tick'' and ``no tick'', respectively.

For a quantitative analysis, it is convenient to introduce an operator that counts the ticks generated by each of the players. Let  us denote by $\tau^A_j$  the projection operator $\tau^A$ applied to player $A$'s $j^{\mathrm{th}}$ tick register, and define $\nu^A_j = \id_{T_j} - \tau_j^A$.  Furthermore, for any $N \in \mathbb{N}$ and $0 < k_1 < k_2 < \cdots < k_s < N$, we  define the operator 
\begin{align} \label{eq_ticklocations}
  \Pi^N_{k_1, k_2, \ldots, k_{s}}
=  \quad & \nu_1 \cdots \nu_{k_1-1} \tau_{k_1}\\
   \quad  & \nu_{k_1+1} \cdots \nu_{k_2-1} \tau_{k_2} \nonumber \\
   \quad & \cdots   \nonumber\\
    \quad  & \nu_{k_{s-1}+1} \cdots \nu_{k_s-1} \tau_{k_s}   \nonumber
  \end{align}
on $T_1 \otimes \cdots \otimes T_N$. Our measure of success in the Alternate Ticks Game can then be formalised as follows. 

\begin{mydef}  
For two clocks the {\it success probability for $t$ ticks} is defined as 
\begin{align}
  p_t =  \lim_{N \to \infty} \tr(\rho^N_A \otimes \rho^N_B \Pi^N_t) \ ,
\end{align}
where $\{\rho^N_A\}_{N \in \mathbb{N}}$ and $\{\rho^N_B\}_{N \in \mathbb{N}}$ are the time scales of the two clocks and where $\Pi^N_t$ is the projector defined by\footnote{The specific expression captures the case of $t$ even, but can easily be adapted to $t$ odd.}
\begin{align}
  \Pi^N_t =  \! \! \! \! \sum_{\substack{0 < k_1 < k_2 < k_3 < k_4 \\ < \cdots <  k_{t-1} < k_t < N}} \! \! \! \!  \Pi^{N}_{k_1, k_3, \ldots, k_{t-1}} \otimes   \Pi^{N}_{k_2, k_4, \ldots, k_t }  \ ,
\end{align}
with the projectors in the sum given by~\eqref{eq_ticklocations}. 
\end{mydef}

Operationally,  $p_t$ corresponds to the probability that {\it at least} $t$ ticks are produced in the correct {\it alternating} order. In particular, we have $p_{t'} \leq p_{t}$ whenever $t' \geq t$. There are different ways to condense this probabilistic statement into a single quantity.  One would be to consider the {\it expected} (or the average) number $\bar{t}$ of ticks until the referee detects a failure, i.e., 
\begin{align}
   \bar{t} = \sum_{t} p_t t \ .
\end{align}
This is the figure of merit that we will adopt in Section~\ref{sec:Results}. Alternatively, one may ask for the maximum number $t^\delta_{\max}$ of ticks such that the failure probability is below a specified {\it threshold} $\delta$, i.e., 
\begin{align*}
  t^\delta_{\max} = \max\{t: p_t \geq 1-\delta \} \ .
\end{align*}

We observe that two players equipped with the NOT-based clock described in Section~\ref{sec:Time} could continue the game arbitrarily, i.e., $p_t = 1$ for all $t \in \mathbb{N}$. In this sense, the NOT-based clock is perfectly precise. However, as already mentioned earlier, it is not continuous. Conversely, the $\epsilon$-continuous clock described in Section~\ref{sec:Cont} has a probabilistic behaviour, which would at some point lead to a failure of the alternating ticks condition. However, for large sizes $d \gg 1$ of the clockwork, this failure may happen only after many tick signals. In fact, if the size is unbounded, there may be strategies that achieve an arbitrarily large expected number~$\bar{t}$ of ticks. 
  
 \section{Two specific strategies for the Alternate Ticks Game} 
   \label{sec:Results}
    
In this section we describe two specific strategies for the Alternate Ticks Game and present selected numerical results illustrating their performance. The strategies are defined by the choice of quantum clocks by the two players, $A$ and $B$. Specifically, we consider clocks of the form~\eqref{eq_typicalclock} characterised by the following parameters: 
\begin{enumerate}
	\item the dimension $d$ of the clockwork $C$ and its initial state $\rho_C^{0}$;
	\item a unitary map $\cM^{\mathrm{int}}(\rho): \rho\to {\rm e}^{-{\rm i}  H^{{\rm int}} \theta}\,\rho\, {\rm e}^{{\rm i}  H^{{\rm int}} \theta}$ specified by a Hermitian operator $H^{{\rm int}}$ as well as a parameter $\theta$;
	\item a map $\cM^{\mathrm{meas}}$ of the form~\eqref{eq_meas_reset}, specified by nonnegative operators  $\pi_0$ and $\pi_1$ and a state $\hat{\rho}_C$. 
\end{enumerate}

We assume that the two players use identical quantum clocks, except for their initial states, which are chosen as
\begin{equation}
	\rho_{C}^{0,A}=\proj{0}, \quad \rho_{C}^{0,B}=\proj{\left\lfloor {\textstyle \frac{d}{2}} \right\rfloor},
\end{equation}
where $\{\ket{0}, \ket{1},\ldots, \ket{d-1}\}$ is an orthonormal basis for the clockwork $C$. 

The first specific strategy is inspired by Peres' model of a simple quantum clock~\cite{Peres} and defined by the Hermitian operator
\begin{subequations}
\begin{align}
	H^{\mathrm{int}}_P={\rm i}\ln U_P\ ,
\end{align}
where 
\begin{equation}
	U_P=\ketbra{0}{d-1} + \sum_{k=0}^{d-2} \ketbra{k+1}{k}
\end{equation}
\end{subequations}
is a unitary matrix that cyclically permutes the basis states --- bringing state $\ket{k}$ to $\ket{k+1\,\text{mod}\, d}$ for all $k \in\{0,\ldots,d-1\}$. The second strategy we consider is based on a clock defined by
\begin{align}
	H^{\mathrm{int}}_{S}=U_P + U_P^\dag\ .
\end{align}
For both types of clock the map $\cM^{\mathrm{meas}}$ is defined by the state $\hat{\rho}_C = \proj{0}$ and the  operators
\begin{equation}\label{Eq:POVMTick}
  \pi_0 = \id-\pi_1 \quad \text{and} \quad \pi_1 =  \delta\sum_{j=d-d_0}^{d-1} \proj{j} \ ,
\end{equation}
where we chose $d_0=\lceil\frac{d}{10} \rceil$.  

\begin{figure}[h]
\includegraphics[scale=0.45]{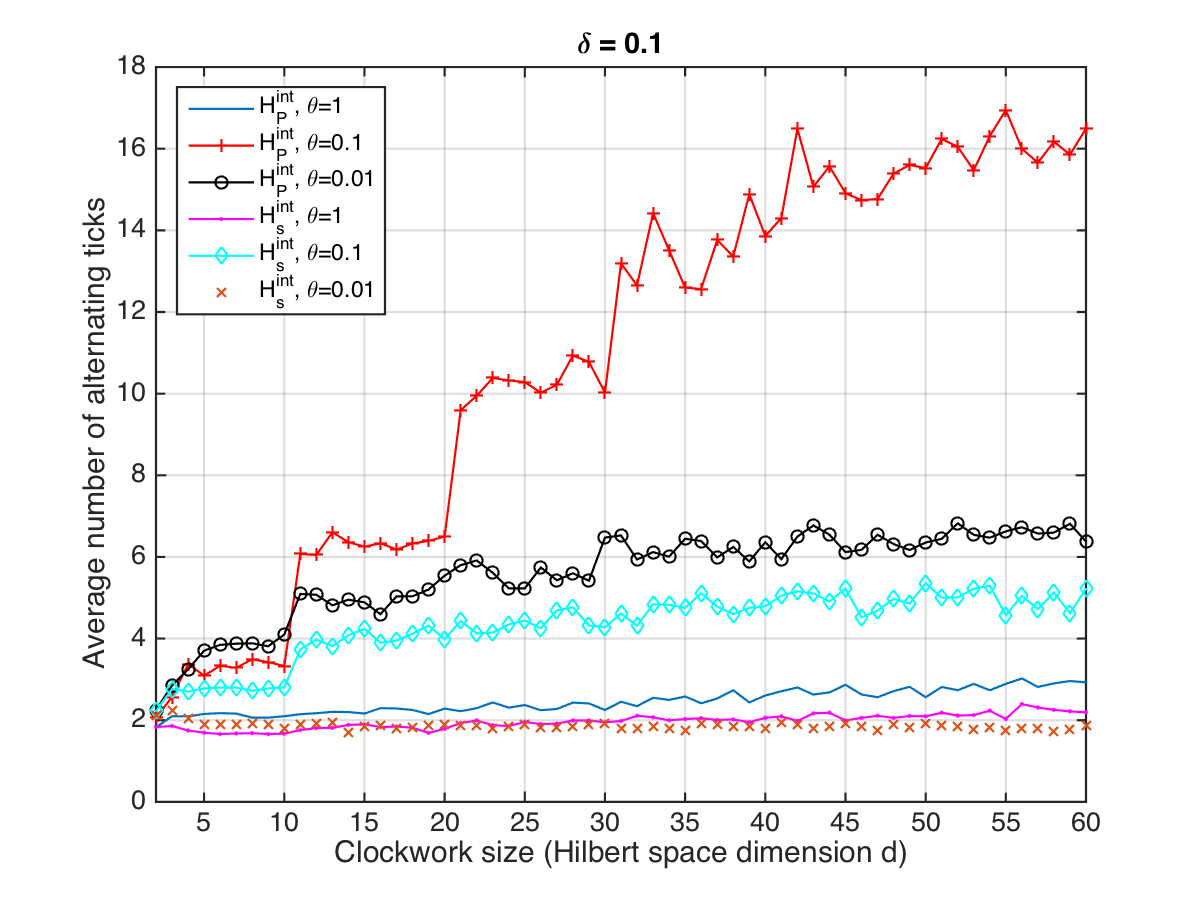} 
\caption{{\it Performance in the Alternate Ticks Game for the two strategies described in the text,  fixed $\delta$, and varying~$\theta$.} The plot shows the average number of alternating ticks as a function of  the size of the clockwork, $d\in\{ 2,...,60\}$, for different step sizes, $\theta \in \{1, 0.1, 0.01\}$ (with $\delta=0.1$). Each data point plotted is the averaged result of $500$ runs of the simulation.}
\label{Fig:Delta0.01}
\end{figure}

\begin{figure}[h]
\includegraphics[scale=0.45]{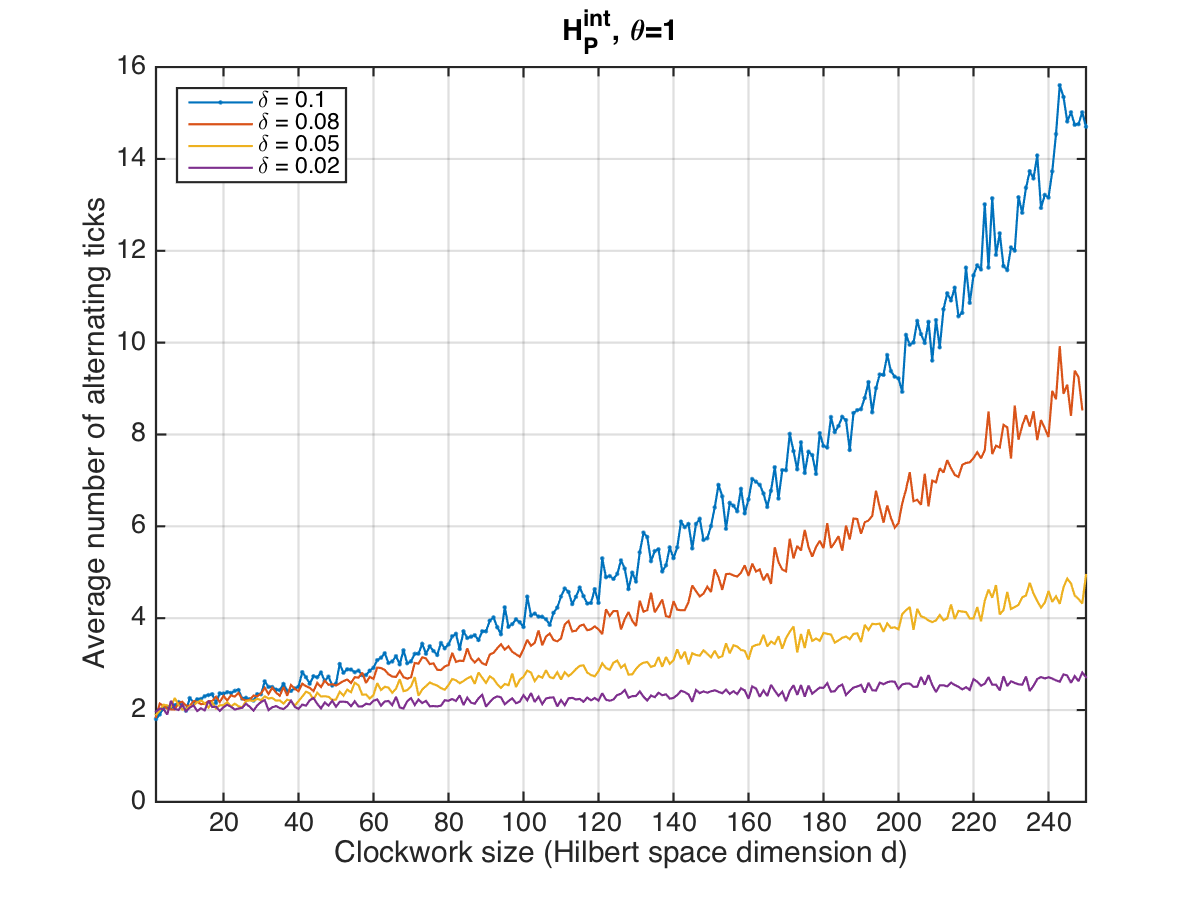} 
\caption{{\it Performance in the Alternate Ticks Game for the strategy defined by the Peres model, fixed~$\theta$, and varying~$\delta$.} The plot shows the average number of alternating ticks as a function of the size of the clockwork,  $d\in\{2,\ldots,250\}$, for different choices of~$\delta\in\{0.02, 0.05, 0.08, 0.1\}$ (with $\theta=1$). Each data point plotted is the averaged result of $500$ runs of the simulation.}
\label{Fig:theta1}
\end{figure}

The results of our numerical analysis of these strategies in the Alternate Ticks Game are summarised in Figs.~\ref{Fig:Delta0.01} and~\ref{Fig:theta1}. The performance obviously depends strongly on the choice of $H^{\mathrm{int}}$ as well as the parameters $\theta$ (which determines the step size in between two interactions of the clockwork with the tick register) and $\delta$ (which determines the strength of the interaction with the tick register). Nonetheless, the results clearly show that the number of achievable alternate ticks increases with the size~$d$ of the clock, at least in the regimes that we explored.

While the optimal choice of $\theta$ depends  on the size of the clockwork, a quantum clock that evolves according to $H^{\mathrm{int}}_P$ with $\theta=1$ has a good performance for a large range of clock sizes, as shown in Fig.~\ref{Fig:theta1}. This choice of parameters therefore has the potential of reaching arbitrary precision in the limit of large clockwork sizes. We expect that this remains true even if one imposes stronger continuity  by further decreasing the value of the parameter~$\delta$.

Although the particular quantum clocks we considered here are already reasonably precise (in terms of the number of alternating ticks generated) they are certainly not optimal. While the construction of good quantum clocks appears to be a non-trivial task, it is certainly an interesting and important one. In fact, the  question of what features make good operational clocks is still largely open.

 \section{Conclusions and future work}
 \label{sec:Conclusion}
  
The aim of this work is to propose a framework to study local clocks and their synchronisation in quantum mechanics.  The Alternate Ticks Game is a  simple and natural approach to quantifying the extent to which quantum clocks can be synchronised  in an operationally meaningful manner. Considering the performance of two copies of a clock in the game, we also obtain a measure for the accuracy of a single quantum clock. Our numerical results show that, even with small-size clocks, a reasonable performance is reachable.

Nonetheless, for any given size of the clock --- measured by the dimension of the state space of the clockwork --- there must exist a non-trivial upper bound on the maximum number of alternating ticks achievable, irrespective of the strategy adopted by the players. This is due to the unavoidable interaction between the clockwork and its environment (the tick registers in our model) and the nature of quantum mechanics --- interactions degrade the ideal evolution of the system and measurements introduce disturbances. We note that, for a different notion of quantum clocks introduced by Salecker and Wigner~\cite{Salecker1958}, the limitations of time measurements have also been discussed, yet from a different perspective, considering the minimal mass required by the clock system (see also~\cite{Ng2008}). 

The related question of uncertainty in time measurements has a long history. The famous uncertainty relations by Heisenberg, Robertson and others~\cite{Heisenberg, Robertson} were derived for  observables in quantum theory. Since coordinate time is not such an observable,  the analogous time-energy uncertainty relations are only applicable in specific cases, e.g., when considering the minimal time needed for a particle to be excited from one energy state to another~\cite{Busch}. Similarly, entropic uncertainty relations~\cite{Deutsch,Uffink,BertaM} can only be used for quantum observables, and are hence not applicable to  coordinate time.  However, it is conceivable that such relations can be obtained for clock time. The operational approach presented here may serve as a starting point towards this goal. 

The synchronisation of clocks is a widely studied problem and commonly employed approaches include  phase estimation, light signals exchanged between parties, or shared entanglement, as described, e.g., in~\cite{Chuang2000,Jozsa2000,SethLl,Preskill2000,Spekkens}. It was found, however, that shared entangled states provided by a third party do not seem to increase the performance of these synchronisation techniques ~\cite{Preskill2000,Spekkens}. One may therefore suspect that the use of entanglement would also not increase the performance of two players in the Alternate Ticks Game.   

There are various possible directions of further research. Clearly, the game that we have introduced  can be generalised to one involving multiple players, i.e., multiple local clocks. One may then impose the synchronisation criterion  that no two successive ticks should come from the same player, or other variations thereof. Conceivably, by increasing the number of players, one can obtain a combined time scale with  smaller intervals (as measured by some external time parameter) between consecutive ticks  --- corresponding to one that would be produced by a clock with higher precision. Such  a construction may also be used to define a  global but still operational notion of time in non-relativistic quantum mechanics. Conversely, it may be worth investigating the impact of relativistic effects to the operational notion of time considered here.

Finally, we briefly comment on the connection between our work and some less closely related research areas. Over the past years there has been renewed interest in the question of how quantum mechanics can be reconciled with general relativity (GR) --- one of the biggest problems of modern physics. Here the ``problem of time" arises as witnessed by the apparent impossibility to combine the  time parameter used to describe the evolution of  quantum systems with the structure of GR, where time is defined locally. This has led to interesting novel approaches to define time. For example, since  the Wheeler-DeWitt equation~\cite{DeWitt} corresponds to a timeless Universe, a possible way to regard time as correlations among systems~\cite{Peres, Wootters}. Our approach has various similarities to this idea. In particular, our notion of time scales is ultimately a stationary concept and the question of synchronisation may be reformulated as the question of whether the correlations between two time scales have a certain desired structure.

\section*{Acknowledgments}

The authors would like to thank \"Amin Baumeler, Roger Colbeck, Paul Erker, Cisco Gooding, Mauro Iazzi, Philipp Kammerlander, Chiara Marletto, Sandu Popescu, Nicolas Sangouard, Ralph Silva, Andr\'e Stefanov, Vlatko Vedral and Stefan Wolf for discussions and useful comments. This work was supported by the Swiss National Science Foundations through the National Centre of Competence in Research ``QSIT'', the European Commission through the ERC grant No.~258932, as well as by the Ministry of Education, Taiwan, R.O.C., through ``The Aim for the Top University Project'' granted to the National Cheng Kung University (NCKU).
\bibliographystyle{unsrt}

\fontsize{3pt}{1pt}\selectfont
\bibliography{references3}

\end{document}